\documentclass[notitlepage]{revtex4-1}

\usepackage{xr}
\usepackage[dvips]{graphicx}
\usepackage{amssymb,amsfonts,amsmath}
\usepackage{natbib}
\usepackage{color}
\usepackage{subfigure}

\def\half{{\textstyle \frac{1}{2}}}

\def\Tr#1{{\rm Tr}\left( #1 \right)}
\def\Det#1{{\rm Det}\!\left( #1 \right)}
\newcommand{\slfrac}[2]{\left.#1\middle/#2\right.}

\begin{document}


\title{2D and Axisymmetric Incompressible Elastic Green's Functions }


\author{John S Biggins}
\affiliation{Cavendish Laboratory, University of Cambridge, Cambridge, United Kingdom}

\author{Z Wei}
\affiliation{School of Engineering and Applied Science, Harvard University, Cambridge, USA}

\author{L Mahadevan}
\affiliation{School of Engineering and Applied Science, Harvard University, Cambridge, USA}


\date{\today}

\begin{abstract}
\end{abstract}

\begin{abstract}We compile a list of 2-D and axisymmetric Green's functions for isotropic full and half spaces, to complement our letter \emph{Linear elasticity of incompressible solids}. We also extend the isotropic exactly incompressible linear theory from our letter to include isotropic neo-Hookean solids subject to a large pre-strain, and present Green's functions in these cases. The Green's function for a pre-strained half-space reproduces the Biot instability. \end{abstract}
\maketitle

The traditional approach to incompressible linear elasticity only enforces volume preservation to linear order, requiring that the divergence of the displacement be zero. In our letter \emph{Linear elasticity of incompressible solids} we demonstrate one can do better by building a linear theory of elasticity with exact volume conservation in either two-dimensional or axisymmetric situations. We also assess the merits of the exact approach by considering explicitly the response of an incompressible 2D medium to a point force. As we discuss in our letter, in 2D or axisymmetric situations, traditional elasticity solutions are described by a stream-line function and a pressure field, both of which are functions of the reference state coordinates. The solution to the corresponding exactly volume preserving problem is the same pair of functions, but now they are functions of one reference coordinate and one target coordinate. Thus solutions from traditional linear elasticity can easily be converted into the exact framework and vice-versa. 

Our purpose here is to document in one place all the main point-force solutions for incompressible 2D and axisymmetric elastic bodies. We express these solutions in the new exactly volume preserving framework but, as noted above, a trivial substitution will transform them into their traditional counterparts. We also extend the isotropic exactly incompressible theory to encompass the linearized response of an isotropic neo-Hookean material with a large pre-strain, and document point-force responses in this case. Many of the responses included in this manuscript have been calculated before within the traditional elastic framework, but we believe there is still considerable value in drawing them all together and expressing them in one language. This document is fairly discursive and should generally permit the reader to construct the solutions themselves. However, some of the solutions, particularly those involving pre-strains and half-spaces, are algebraically cumbersome, so we also provide with this document two mathematica notebooks that readers can use to automatically verify the  the claimed solutions.

\subsection{Incrementally linear elasticity for planar systems with large pre-strains}
In many situations associated with the deformation of soft materials, there are large homogeneous residual strains associated with polymerization, growth, swelling and shrinkage. Since many strictly incompressible systems are rather soft, such large deformations are  easily achieved, and linearizing around a pre-strained state greatly increases the scope of the linear theory. Such calculations are likely to be required when evaluating the stability of a highly deformed state as, for example, in Biot's celebrated compressive instability \citep{biotbook, hohlfeld2011} or the onset of director rotation in liquid-crystal elastomers \citep{bigginssemisoft}.  We include an elastic pre-strain by taking our elastic reference state (with coordinates $(p,q)$) as already having undergone a large homogeneous deformation of the form $F_0=\mathrm{diag}(\lambda,1/\lambda)$, then undergoing an additional small displacement $\mathbf{u}(p,q)$ that leads to the additional deformation gradient $F_1$, so that the  the total deformation gradient (from the undeformed state) is
\begin{equation}
F=F_1\cdot F_0.
\end{equation}
The effective energy is then
\begin{equation}
\tilde{E}=\half \mu \Tr{F_1\cdot F_0\cdot F_0^T \cdot F_1^{T}}+P(\Det{F_1}-1)-\mathbf{f}\cdot \mathbf{u}.
\end{equation}
Minimizing this with respect to $\mathbf{u}$ and $P$ gives
\begin{equation}
\nabla \cdot (\mu F_1\cdot F_0\cdot{F_0}^{T}-P F_1^{-T})=-\mathbf{f} \hspace{3 em} \Det{F_1}=1,\label{tradnleqns}
\end{equation}
where, the divergence is taken in the $p-q$ state. As in our letter, we then implement the constraint $\Det{F_1}=1$ exactly by parameterising all the quantities in our problem via the mixed coordinates $x$ and $q$.  We can then use the function $\psi(x,q)$ to describe $F_1(x,q)$, via eqn.\ (13) from our letter. To linearize about the pre-strained state we write $\psi(x,q)=xq+\alpha(x,q)$ and $P=\mu(P_0+\kappa(x,q))$, where $P_0$ is a constant (possibly large) pressure associated with the pre-strain, and, in the linear regime, we expect both $\kappa$ and $\alpha$ to be small. We then expand the above equation of equilibrium to first order in $\alpha$ and $\kappa$, noting, as in out letter,  that, to linear order, the partial derivative identities  $\frac{\partial}{\partial x}\big|_z=\frac{\partial}{\partial x}\big|_q$ and $\frac{\partial}{\partial z}\big|_x=\frac{\partial}{\partial q}\big|_x$ hold, and  once again using use the linearized forms of $F_1$ and $F_1^{-T}$ (eqns (15-16) from our letter). Linearization yields
\begin{equation}
\frac{1}{\lambda^2}\left(
 \begin{array}{c}
-\alpha _{{qqq}}-\lambda^4 \alpha _{{xxq}} \\
\alpha _{{xqq}}+\lambda^4 \alpha _{{xxx}}
\end{array}
\right)-\left(
\begin{array}{c}
 \kappa_x \\
\kappa_q
\end{array}
\right)=-\mathbf{f}/\mu.\label{2deqilib}
\end{equation}
In a region with no external force ($\mathbf{f}=0$)  we can eliminating  $\kappa$ to find that $\alpha(x,q)$ satisfies an anisotropic analog of the biharmonic equation,
\begin{equation}
\alpha _{{qqqq}}+\left(1+\lambda^4\right) \alpha _{{xxqq}}+\lambda^4 \alpha _{{xxxx}}=0,
\end{equation}
(derived for swollen systems by \cite{ben2010swelling}), which we can factorize as
\begin{equation}
\left(\frac{\partial^2}{\partial x^2}+\frac{\partial^2}{\partial q^2}\right)\left(\frac{\partial^2}{\partial x^2}+\lambda^4 \frac{\partial^2}{\partial q^2}\right)\alpha=0.\label{mastereqn2d}
\end{equation}
In the case of no pre-stretch ($\lambda=1$) this reduces to the bi-harmonic discussed in our letter.

\subsection{Green's functions for unstrained full space}\label{fullunstrained}
In two dimensions we expect the stress (and strain) associated with a point force to vary inversely as the distance from its point of application, and hence the displacement to vary logarithmically with this distance. This leads us to try the form
\begin{equation}
\alpha(x,q)=A x \log \left(q^2+x^2\right)+B q \log \left(q^2+x^2\right),
\end{equation}
a function that is biharmonic. Substituting this form into the equation of equilibrium (eqn.\ \ref{2deqilib}) we see that the incremental pressure must be given by
\begin{equation}
\kappa(x,q)=\frac{4 (Bx-Aq)}{q^2+x^2}.
\end{equation}
This $\alpha$ and $\kappa$ satisfy  the equations of elasticity (eqn.\ \ref{2deqilib}) with a diverging stress at $x=q=0$, the point of application of the force. To find the force's magnitude we imagine cutting out the infinite strip of material $|x|<a$ with the surface normals $(\pm1,0)$. The stress tensor from eqn.\ (\ref{tradnleqns}), $\sigma=\mu F_1\cdot F_0\cdot{F_0}^{T}-P F_1^{-T})$, relates normals in the $p-q$ space to forces in $x-z$ space. For a space without pre-strain it reduces to the  standard PK1 stress $\mu F-P F^{-T}$. Expanding $\sigma\cdot(\pm1,0)$ to first order  and integrating around the strip yields the force
\begin{equation}
\int_{-\infty}^{\infty} \sigma(a,q)\cdot(1,0)  \mathrm{d}q+\int_{\infty}^{-\infty} \sigma(-a,q)\cdot(-1,0)  \mathrm{d}q=\left(
\begin{array}{c}
 -8 B \pi  \mu  \\
 8 A \pi  \mu 
\end{array}
\right).
\end{equation}
Setting the component of the force in the $q$ direction to $f^{(q)}$, and the component of the force in the $x$ direction to $f^{(x)}$, the full space Green's function is 
\begin{equation}
\alpha(x,q)=\frac{(f^{(x)} q-f^{(q)} x) \log\left(q^2+x^2\right)}{8 \pi  \mu }\mathrm{\ \ \ \ and \ \ \ \ } \kappa(x,q)=\frac{f^{(q)} q+f^{(x)} x}{2 \pi   \mu(q^2+x^2) } \label{2dsol}.
\end{equation}

\subsection{Planar Green's functions for pre-strained full space}
\begin{figure}\centering
\subfigure[$\alpha_2$]{\label{onecompfuncs}\includegraphics[width=0.24\textwidth]{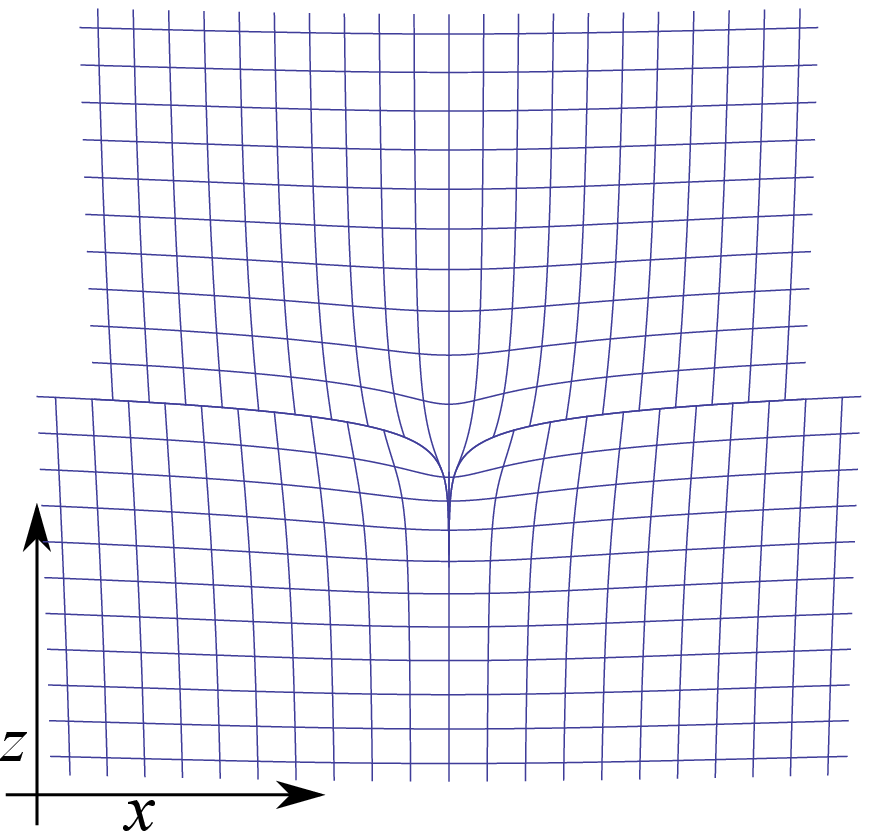}}
\subfigure[$\lambda=0.6$]{\label{streamline}\includegraphics[width=0.24\textwidth]{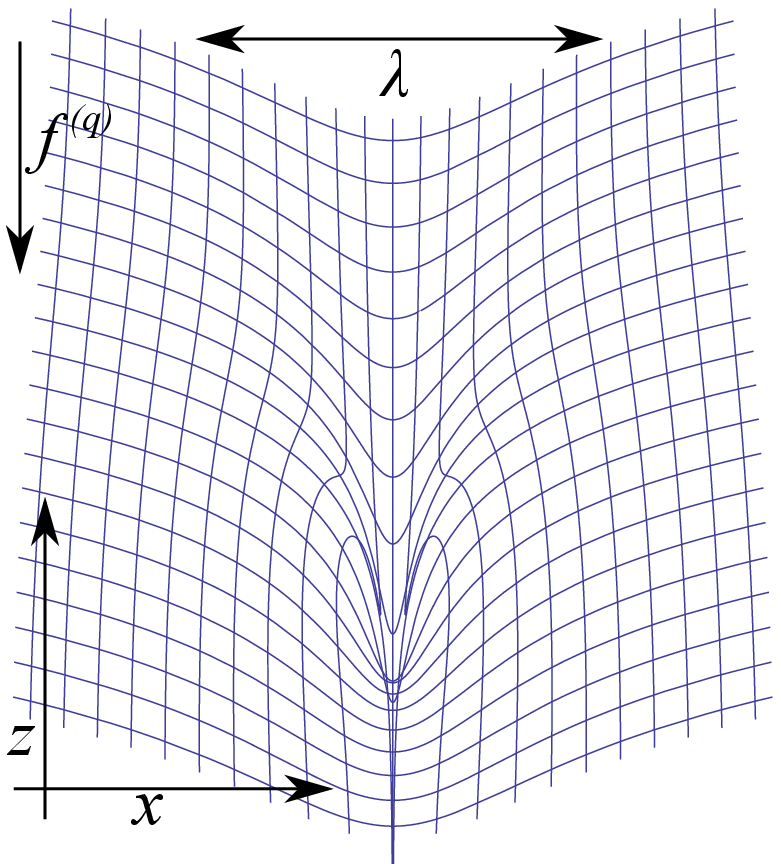}}
\subfigure[$\lambda=1.2$]{\label{linear}\includegraphics[width=0.24\textwidth]{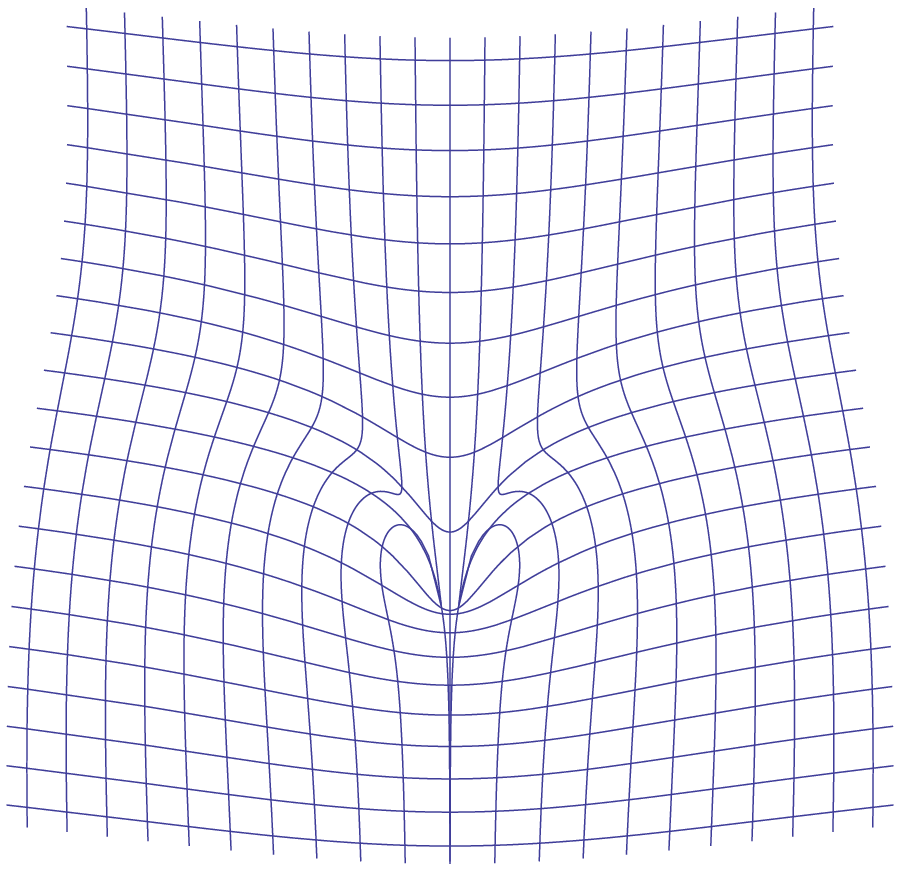}}
\subfigure[Conventional, $\lambda=0.6$]{\label{linear}\includegraphics[width=0.24\textwidth]{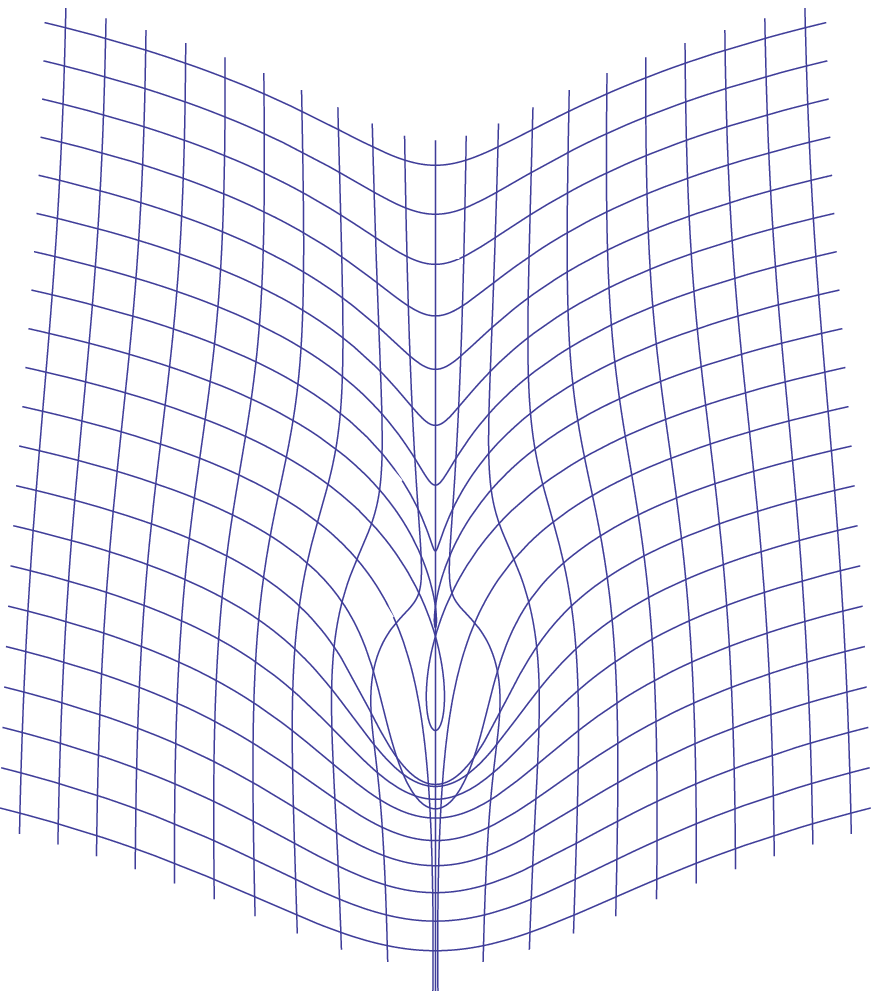}}
\caption{(a) A plot of the deformation field associated with harmonic function $\alpha_2$ (eqn. (\ref{al2})) showing promising point-force like behavior at the origin but also a discontinuity in displacement (corresponding to a dislocation) along the line $q=0$. Subfigures (b) and (c):  Deformation caused by a point force of magnitude $2 \mu$ in two dimensions after imposition of a large pre-strain (eqn.\ (\ref{alphav}))  calculated using exact area preservation. The solutions are formed by taking superpositions of the fields $\alpha_2$ and $\alpha_4$ to cancel out their dislocations. The grids shown were square grids in the compressed base state with lattice-spacing 0.1, and the values of $\lambda$ given are the compression ratios of the pre-strain in the direction perpendicular to the direction of the force. Compression transverse to the direction of application of the force leads to substantially enhanced response. Subfigure (d): same as (b) but the traditional rather than area preserving solution.}\label{2dkelvincompressed}
\end{figure}

The pre-strain breaks the symmetry between the two material directions making our material and hence our Green's functions anisotropic. Indeed, eqn. (\ref{mastereqn2d}) is analogous to the equation governing two dimensional transversely isotropic elastic systems \citep{ting1996anisotropic}.  We note that the above equation for $\alpha$ factorizes into a pair of commuting operators, a Laplacian and a scaled Laplacian, so that we expect to see our Green's functions to be a combination of harmonic  or  scaled harmonic functions.  Scaling the unstrained case suggests functions of the form $x \log(x^2+q^2)$ and $q \log(x^2+q^2)$, but these are not harmonic. However, by taking the  imaginary parts of $(x+ i q)\log(x+ i q)$ and $(i x+  q)\log(i x+  q)$ which are harmonic, we write 
\begin{align}
\alpha_1(x,q) &=\half q \log(x^2+q^2) +x \arctan(q/x)\label{al1}\\
\alpha_2(x,q) &=\half x \log(x^2+q^2) +q \arctan(x/q)\label{al2}.
\end{align}
and further define the analogous functions for the scaled Laplacian
\begin{align}
\alpha_3(x,q) &=\half q \lambda^2 \log(x^2+\lambda ^4 q^2) +x \arctan(\lambda^2 q/x)\label{al3}\\
\alpha_4(x,q) &=\half x \log(x^2+\lambda^4 q^2) +q\lambda^2 \arctan(x/(\lambda^2 q))\label{al4}.
\end{align}
These functional forms all diverge at the origin with the correct scaling for a point force and obey eqn.\ (\ref{mastereqn2d}). Substituting these forms into (\ref{2deqilib}) allows us  to deduce the pressure fields:
\begin{align}
\kappa_1(x,q)=\frac{ x \left(1-\text{$\lambda $}^4\right)}{\left(q^2+x^2\right) \text{$\lambda $}^2} \mathrm{,\ \ \ \ \ \ } \kappa_2(x,q)&=\frac{  q \left(1-\text{$\lambda $}^4\right)}{\left(q^2+x^2\right) \text{$\lambda $}^2}\label{k2}\\
\kappa_3(x,q)=0\mathrm{,\ \ \  \ \ \  \ \ \ \ \ \ \ \ \ \ \ \ \ \ \  \ \ }\kappa_4(x,q)&=0.\label{k4}
\end{align}
We now  have four functions all of which have the required properties, but our final answer should only have two degrees of freedom denoting the horizontal and vertical components of the applied force. The reason for this conundrum is that the functions above which are directly analogous to the ``one component" Green's functions in anisotropic elasticity \citep{ting1996anisotropic} give rise to displacement fields with dislocations running through the origin that extend to infinity. They are associated with the behavior of $\arctan$ functions that appear in the expressions (\ref{al1})-(\ref{al4}) when the denominator of the argument passes through zero. These can clearly be seen in a plot of the associated displacement field shown in fig. \ref{onecompfuncs}. To cancel out these discontinuities we add linear combinations of the two functions with the discontinuities in the same place, which allow us to define a pair of Green's functions for the displacements 
\begin{align}
\alpha^{(q)}(x,q)&\propto 2 q\arctan\left(\frac{x}{q}\right)-2 q\arctan\left(\frac{x}{q \lambda^2}\right)+x \log\left(q^2+x^2\right)- (x/\lambda^2) \log\left(x^2+q^2 \lambda^4\right)\\
\alpha^{(x)}(x,q)&\propto -2 x\arctan\left(\frac{q}{x}\right)+2 x\arctan\left(\frac{q \lambda^2}{x}\right)-q \log\left(q^2+x^2\right)+q \lambda^2 \log\left(x^2+q^2 \lambda^4\right).\end{align}
These functions do now give rise to continuous strain fields, and correspond to Green's functions for a horizontal $(x)$ and vertical $(q)$ point force respectively. Normalizing these functions by requiring the $\lambda \to1$ limit matches the isotropic result yields
\begin{align}
\alpha^{(q)}(x,q)&=\frac{f^{(q)} \lambda^2 \left(2 q\arctan\left(\frac{x}{q}\right)-2 q\arctan\left(\frac{x}{q \lambda^2}\right)+x \log\left(q^2+x^2\right)- (x/\lambda^2) \log\left(x^2+q^2 \lambda^4\right)\right)}{4 \pi \mu \left(1-\lambda^4\right)}\label{alphav}\\
\alpha^{(x)}(x,q)&=\frac{f^{(x)}\lambda^2 \left(2 x\arctan\left(\frac{q}{x}\right)-2 x\arctan\left(\frac{q \lambda^2}{x}\right)+q \log\left(q^2+x^2\right)-q \lambda^2 \log\left(x^2+q^2 \lambda^4\right)\right)}{4 \pi \mu  \left(1-\lambda^4\right)}.
\end{align}
The associated pressure fields are given by:
\begin{equation}
\kappa^{(q)}(x,q)=\frac{f^{(q)} q}{2 \pi \mu   \left(q^2+x^2\right)}\mathrm{,\ \ \ \ \ and \ \ \ \ \ } \kappa^{(x)}(x,q)=\frac{f^{(x)} x}{2 \pi \mu  \left(q^2+x^2\right)}.\label{kappav}
\end{equation}
Plots of these Green's functions are shown in fig. \ref{2dkelvincompressed} for two different values of the pre-strains and show that for $\lambda<1$ corresponding to horizontal pre-compression transverse to a vertical force leads to a larger displacement response than when $\lambda>1$ corresponding to a horizontal pre-stretch and a  vertical force. The traditional linear elastic Green's function for a point force, found for in any prestrined initially isotropic hyper-elastic 2-D solid was found, using a traditional stream-line function, by \cite{bigoni2002green} and is plotted in fig.\ \ref{2dkelvincompressed}d. As expected, in regions of high strain, it suffers severe area changes, whereas our function does not. 

\subsection{Planar Green's functions for pre-strained half spaces}
We now consider the Green's functions for a force acting in the elastic half space $q<0$ and construct them from the full space Green's functions via the superposition of image solutions outside of the elastic half space to satisfy the boundary conditions on the free surface. An elastic half-space has a free surface ($q=0$) which must be stress free. From eqn.\ (\ref{tradnleqns}), we see that in our case this requires 
\begin{equation}
(\mu F_1\cdot F_0\cdot{F_0}^{T}-P F_1^{-T})\cdot(0,1)=(0,0).
\end{equation}
Although this expression follows directly from eqn.\ (\ref{tradnleqns}), some readers may have been expecting it to involve the PK1 stress tensor, $\mu F_1\cdot F_0-P F_1^{-T})F_0^{-T}$, which relates reference state normals to final state forces. It does not because the unit normal in question is defined in our pre-strained elastic reference state, not the zero strain state. However, since in our case the normal is $(0,1)$ in both states (as the pre-strain does not rotate the free surface), one could in-fact use PK1 in the above expression without error. Linearizing the above in the incremental pressure $\kappa$ and the displacement potential $\alpha$  yields
\begin{equation}
\left(
\begin{array}{c}
 -  \alpha _{{qq}}+P_0 \lambda^2 \alpha _{{xx}} \\
 1-P_0 \lambda^2-\kappa \lambda^2+\left(1+P_0 \lambda^2 \right) \alpha _{{xq}}
\end{array}
\right)_{q=0}=\left(
\begin{array}{c}
 0\\0
\end{array}
\right).\label{resstressboundary}
\end{equation}
Considering the case where there is no applied force, whence $\kappa= \alpha=0$, we see that $P_0=1/\lambda^2$. If $\kappa$ and $\alpha$ are given by eqn.\ (\ref{alphav}) and (\ref{kappav}) for the whole space Green's functions corresponding to a vertical  point force at $q=-d$, the left hand side of  eqn.\ (\ref{resstressboundary}) is
\begin{equation}
\frac{f^{(q)}}{2 \pi  \mu \left(\lambda^4-1\right)}\left(
\begin{array}{c}
 \frac{2 x \lambda^2}{d^2+x^2}-\frac{x \left(1+\lambda^4\right)}{x^2+d^2 \lambda^4} \\
 \frac{d \lambda^2 \left(1+\lambda^4\right)}{d^2+x^2}-\frac{2 d \lambda^{4}}{x^2+d^2 \lambda^4}
\end{array}
\right)\label{stresstocancel}.
\end{equation}
Since this is neither an even nor an odd function of $d$ we cannot use mirror image forces to satisfy the free boundary condition. 
However, a linear combination of four image solutions corresponding $\alpha_2(x,q-d), \alpha_2(x,q-d\lambda^2) $ (eqn.\ \ref{al2} and \ref{k2}) and $\alpha_4(x,q-d), \alpha_4(x,q-d/\lambda^2)$ (eqn.\ (\ref{al4}) and (\ref{k4}))  gives four parameters with which to cancel out the four terms in eqn.\ (\ref{stresstocancel}). We may then write the vertical point force Green's function for a half space as
\begin{align}
\alpha^{(q)}_{hs}(x,q)&=\alpha^{(q)} (x, q+d)+a_1\alpha_2(x,q-d)+a_2\alpha_4(x,q-d)+a_3\alpha_2(x,q-d\lambda^2)+a_4\alpha_4(x,q-d/\lambda^2),\notag \\
\kappa^{(q)}_{hs}(x,q)&=\kappa^{(q)} (x, q+d)+a_1\kappa_2(x,q-d)+a_2\kappa_4(x,q-d)+a_3\kappa_2(x,q-d\lambda^2)+a_4\kappa_4(x,q-d/\lambda^2),\label{halfspacverticalnoniso}
\end{align}
and make the boundary stress-free by setting
\begin{align}
a_1=-\frac{f^{(q)} \lambda^2 \left(1+3 \lambda^2-\lambda^4+\lambda^6\right)}{ 2 \pi \mu  \left(-1+\lambda^2\right)^2 \left(-1+3 \lambda^2+\lambda^4+\lambda^6\right)}\mathrm{,\ }a_3&=\frac{ 2f^{(q)} \lambda^2 \left(1+\lambda^4\right)}{ \pi \mu  \left(-1+\lambda^2\right)^2 \left(1+\lambda^2\right) \left(-1+3 \lambda^2+\lambda^4+\lambda^6\right)}\notag \\ 
a_2=a_1/\lambda^2\mathrm{,\ \ \ \ \ \ \ \ \ \  \ \ \ \ \ \ \ \ \ \ \ \ \ \ \ \ \ \ \ \  \ \ \ \ \ \ \ \ \ \ \ \ \ \ \ \ }a_4&=a_3.\label{a1coef} \end{align}
Although these image charges are associated with line dislocations, they are along lines of constant $q$ and lie completely outside the elastic domain, so they are not physically important.

We calculate the half space Green's function for a horizontal force in exactly the same way, now using a trial function of the form 
\begin{align}
\alpha^{(x)}_{hs}(x,q)&=\alpha^{(x)} (x, q+d)+b_1\alpha_1(x,q-d)+b_2\alpha_3(x,q-d)+b_3\alpha_1(x,q-d\lambda^2)+b_4\alpha_3(x,q-d/\lambda^2)\notag \\
\kappa^{(x)}_{hs}(x,q)&=\kappa^{(x)} (x, q+d)+b_1\kappa_1(x,q-d)+b_2\kappa_3(x,q-d)+b_3\kappa_1(x,q-d\lambda^2)+b_4\kappa_3(x,q-d/\lambda^2)\label{halfspacexgreen}.
\end{align}
where the stress-free boundary condition, eqn.\ (\ref{resstressboundary}),  is satisfied by the following choices of the constants:
\begin{align}
b_1=\frac{f^{(x)} \lambda^2 \left(1+3 \lambda^2-\lambda^4+\lambda^6\right)}{ 2 \pi \mu  \left(-1+\lambda^2\right)^2 \left(-1+3 \lambda^2+\lambda^4+\lambda^6\right)}\mathrm{,\ }b_3&=-\frac{2 f^{(x)} \lambda^4 \left(1+\lambda^4\right)}{\pi \mu  \left(-1+\lambda^2\right)^2 \left(1+\lambda^2\right) \left(-1+3 \lambda^2+\lambda^4+\lambda^6\right)} \notag \\ 
b_2=b_1\mathrm{,\ \ \ \ \ \ \ \ \ \ \ \ \ \ \  \ \ \ \ \ \ \ \ \ \ \ \ \ \ \ \ \  \ \ \ \ \ \ \ \ \ \ \ \ \ \ \ \ }b_4&=b_3/\lambda^2.\end{align}

In this case the line-dislocations associated with the image charges are all along the line $x=0$ and do penetrate the elastic half-space, so to restore the continuity of displacement in the half-space we must add a step function in displacement, by adding $\frac{ f^{(x)} \lambda^2 \left(\lambda^2-1\right) x(2 \theta (x d)-1)}{2 \mu (\lambda^6+\lambda^4+3 \lambda^2-1)}$ to $\alpha(x,q)$ given by eqn.\ (\ref{halfspacexgreen}). 

\subsection{Green's functions for unstrained half space}
\begin{figure}\centering
\subfigure[Vertical  area preserving solution (eqn.\ (\ref{verthalf}))]{\label{linear}\includegraphics[width=0.32\textwidth]{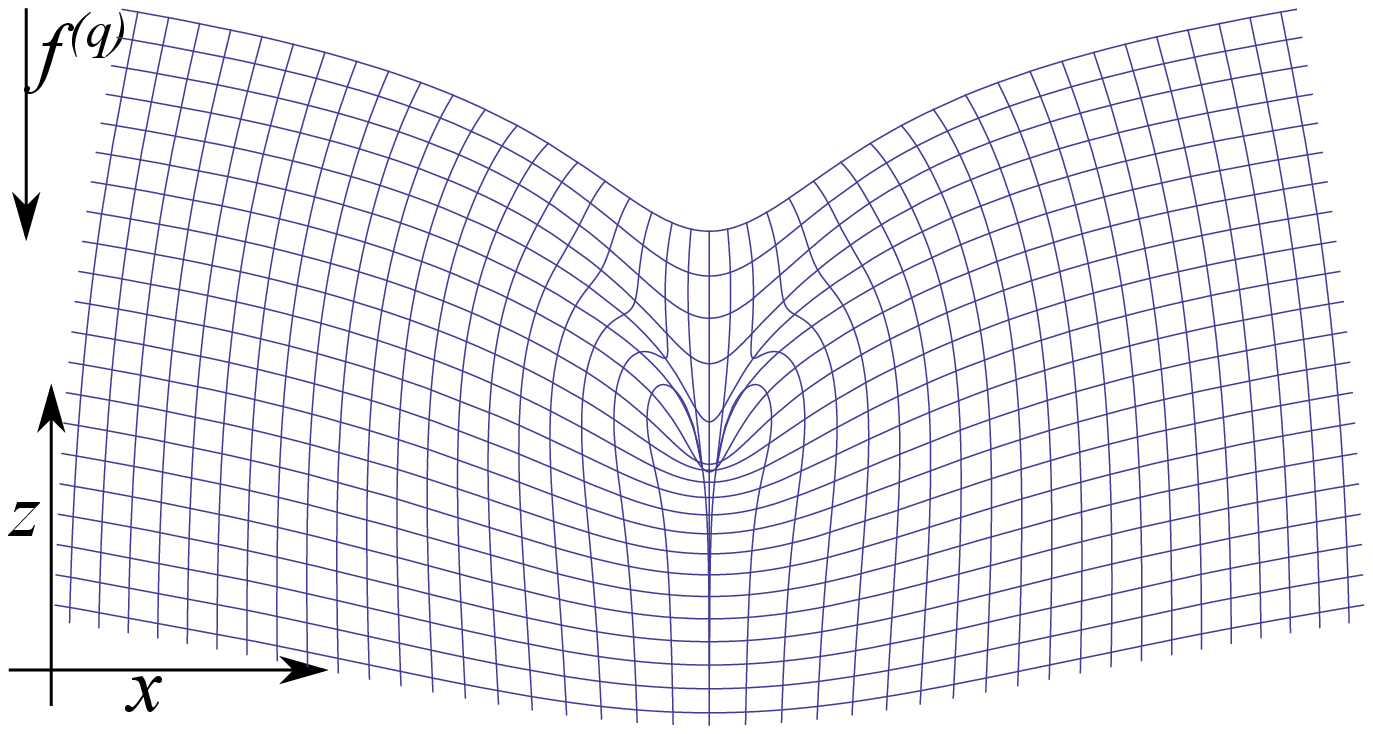}}
\subfigure[Vertical traditional (Melan) solution]{\label{streamline}\includegraphics[width=0.32\textwidth]{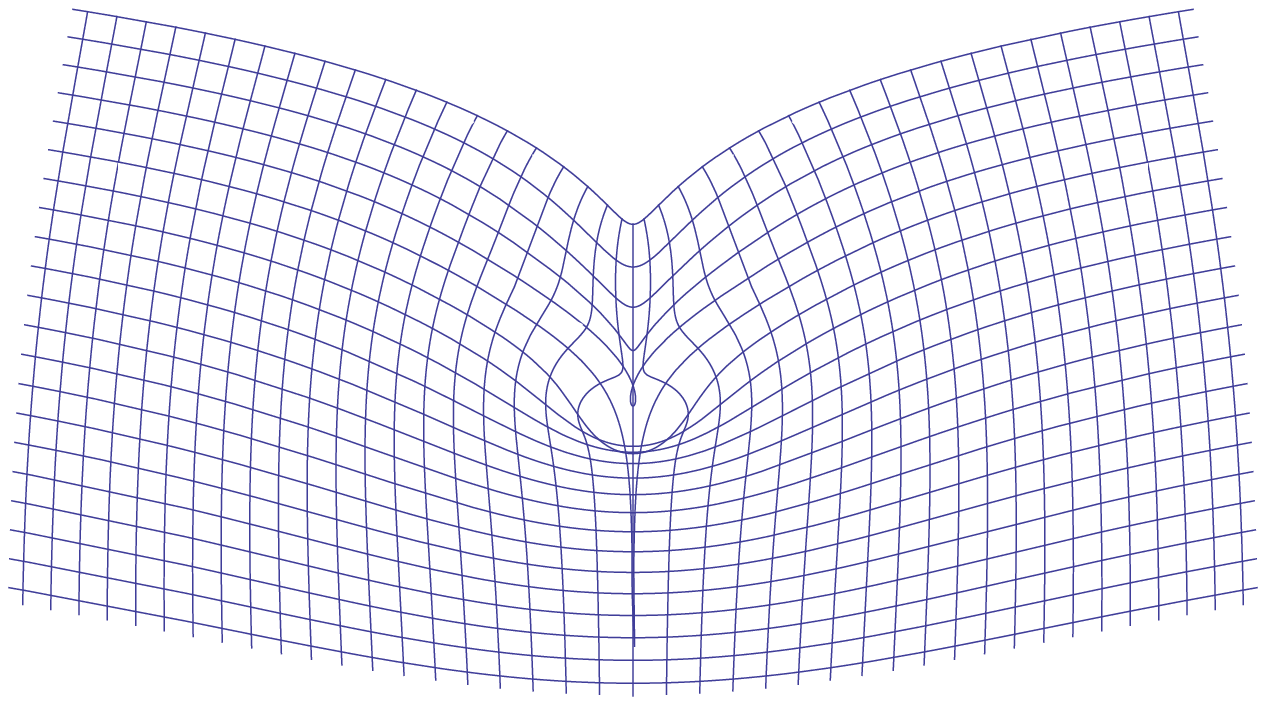}}
\subfigure[Volume changes in traditional vertical solution.]{\label{streamline}\includegraphics[width=0.34\textwidth]{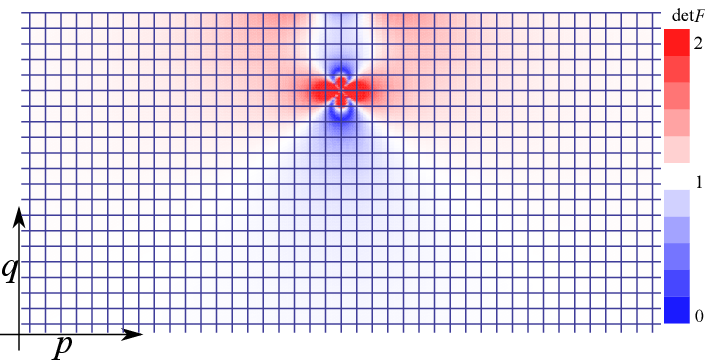}}

\subfigure[Horizontal  area preserving solution  (eqn.\ (\ref{horizhalf}))]{\label{linear}\includegraphics[width=0.32\textwidth]{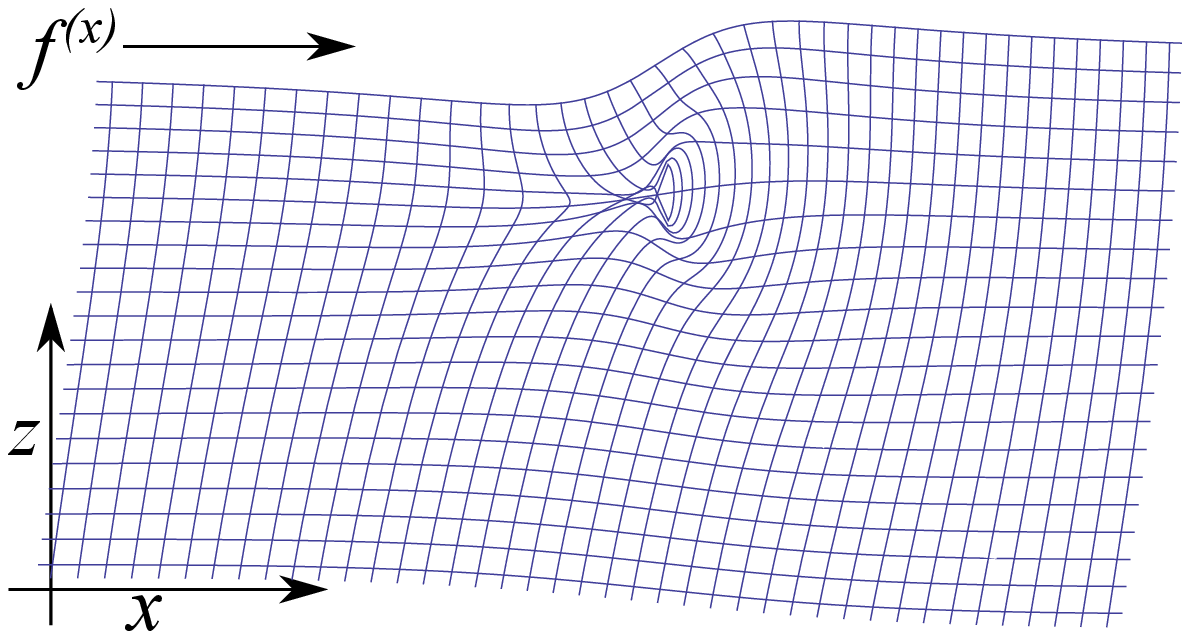}}
\subfigure[Horizontal traditional (Melan) solution]{\label{streamline}\includegraphics[width=0.32\textwidth]{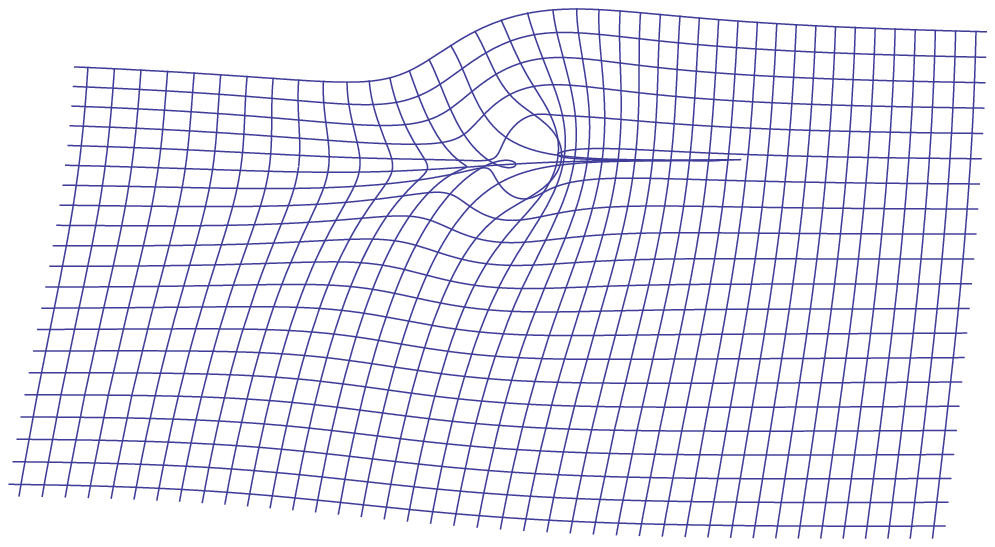}}
\subfigure[Volume changes in traditional horizontal solution.]{\label{streamline}\includegraphics[width=0.34\textwidth]{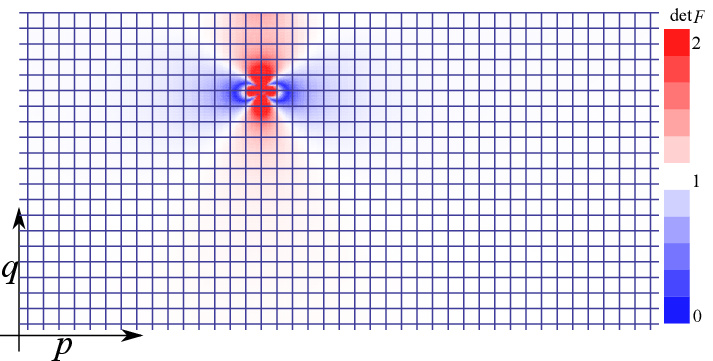}}

\caption{Deformation of a square grid with spacing 0.1 caused by a point force of strength $2 \mu$ applied beneath the surface of an incompressible isotropic half-space at a location $(0,-1/2)$. Figures (a) and (d) are the exactly area preserving responses to forces applied perpendicular and parallel to the surface given by eqns.\ (\ref{verthalf}) and (\ref{horizhalf}), while (b) and (e) are the responses given by the classical linear elastic (Melan) solutions showing severe area distortion near the point of application of the force. Figures (c) and (f) show the reference state for the two cases colored by the volume change caused by the corresponding Melan solutions. }\label{halfspacefig}
\end{figure}

Taking the isotropic limit ($\lambda\to 1$) of eqn.\ (\ref{halfspacverticalnoniso})  gives the isotropic point force Green's function as
\begin{align}
\alpha(x,q)&=-\frac{f^{(q)}  }{8 \pi \mu}\left(\frac{4 d q}{(d-q)^2+x^2}+\log \left(\left((d-q)^2+x^2\right) \left((d+q)^2+x^2\right)\right)\right)\label{verthalf}\\
\kappa(x,q)&=-\frac{f^{(q)}  \left((d-q)^2 (d+q) \left(d^2+2 d q-q^2\right)-2 q \left(2 d^2-d q+q^2\right) x^2-(d+q) x^4\right)}{\pi  \mu  \left((d-q)^2+x^2\right)^2 \left((d+q)^2+x^2\right) }.
\end{align}
for a vertical force at $(0,-d)$ in a half space, while, for a horizontal force we use of eqn. (\ref{halfspacexgreen}) to get
\begin{align}
\alpha(x,q)&=\frac{f^{(x)}}{8 \pi  \mu }\left(\frac{4 \left(d^3-2 d q^2+x^2 (d+q)+q^3\right)}{(d-q)^2+x^2}+(d+q) \log \left(\left((d-q)^2+x^2\right) \left((d+q)^2+x^2\right)\right)\right)\label{horizhalf}\\
\kappa(x,q)&=\frac{f^{(x)} x \left(-d^4-4 d^3 q+4 d^2 q^2+\left(q^2+x^2\right)^2\right)}{\pi  \mu  \left((d-q)^2+x^2\right)^2 \left((d+q)^2+x^2\right)}.
\end{align}
 Once again, this matches the conventional Green's functions for a point force in an elastic half-space \citep{melan1932point} to linear order.  In fig.\ \ref{halfspacefig} we compare this solution to the conventional linear elastic solution and see that while both solutions have divergent displacement and self-intersection in the neighborhood of the force, the conventional solution produces poor area conservation over a wide area, leading to substantially different forms for the free surfaces.

\subsection{Planar higher order and surface Green's functions}
It is straightforward to combine the above Green's functions to construct dipole and quadruple Green's functions with and without moments. For example the Green's function for a horizontal dipole in an uncompressed full two dimensional space is simply
\begin{align}
\alpha(x,q)&=\lim_{a\to 0}\left(\frac{f^{(x)} q \log\left(q^2+(x+a)^2\right)}{8 \pi  \mu }-\frac{f^{(x)} q \log\left(q^2+(x-a)^2\right)}{8 \pi  \mu }\right)=\frac{f^{(x)} q x a}{2 \pi  \left(q^2+x^2\right) \mu }\\
\kappa(x,q)&=\lim_{a\to 0}\left(\frac{f^{(x)} (a+x)}{2 \pi \mu  \left(q^2+(a+x)^2\right)}-\frac{f^{(x)}  (-a+x)}{2 \pi \mu \left(q^2+(-a+x)^2\right)}\right)=\frac{f^{(x)}  \left(q^2-x^2\right) a}{\pi \mu \left(q^2+x^2\right)^2}.\end{align}

We can also look at the limit $d\to0$ to find the Green' s functions for a point force acting on the surface of a half space. 
Since taking these limits is straightforward, we do not explicitly calculate any  examples here but some higher order and surface solutions are shown in tables \ref{fullspacetab} and \ref{surfacetab}.
\begin{table}
\begin{center}
    \begin{tabular}{ |l | l | l |}
    \hline
    Description &$\alpha(x,q)$&$\kappa(x,q)$ \\ \hline
    Isotropic full-space horizontal dipole &$ \frac{ q x}{2 \pi  \left(q^2+x^2\right)   }\notag$ & $\frac{ (q-x) (q+x)}{\pi  \left(q^2+x^2\right)^2}$\\ \hline
    Pre-strained full- Space horizontal dipole & $\frac{ \lambda^2 \left(\tan ^{-1}\left(\frac{q \lambda^2}{x}\right)-\tan ^{-1}\left(\frac{q}{x}\right)\right)}{  \pi  \left(\lambda^4-1\right)}$ &$\frac{ (q-x) (q+x)}{\pi  \left(q^2+x^2\right)^2}$ \\ \hline
Isotropic full-space vertical dipole & $\frac{q x}{2   \pi  \left(q^2+x^2\right)}$ & $\frac{ (q-x) (q+x)}{\pi  \left(q^2+x^2\right)^2}$  \\ \hline
  Pre-strained full-space vertical dipole & $\frac{ \lambda^2 \left(\tan ^{-1}\left(\frac{x}{q}\right)-\tan ^{-1}\left(\frac{x}{q \lambda^2}\right)\right)}{  \pi  \left(\lambda^4-1\right)} $ & $\frac{ (q-x) (q+x)}{\pi  \left(q^2+x^2\right)^2}$ \\ \hline
 Pre-strained full-space horizontal quadrupole & $\frac{q \lambda^2 (-x+q \lambda) (x+q \lambda)}{2   \pi  \left(q^2+x^2\right) \left(1+\lambda^2\right) \left(x^2+q^2 \lambda^4\right)}$& $\frac{ x \left(-3 q^2+x^2\right)}{\pi  \left(q^2+x^2\right)^3}$ \\ \hline
 Pre-strained full-space vertical quadrupole &$\frac{x \lambda^2 (x-q \lambda) (x+q \lambda)}{2   \pi  \left(q^2+x^2\right) \left(1+\lambda^2\right) \left(x^2+q^2 \lambda^4\right)}$&$-\frac{ q \left(q^2-3 x^2\right)}{\pi  \left(q^2+x^2\right)^3}$\\ \hline
    \hline
    \end{tabular}
\end{center}
\caption{Full space higher order Green's functions.}\label{fullspacetab}
\end{table}


\begin{table}
\begin{center}
    \begin{tabular}{ |p{0.18\textwidth} | p{0.59\textwidth} | p{0.24\textwidth} |}
    \hline
    Description &$\alpha(x,q)$&$\kappa(x,q)$ \\ \hline
    Isotropic half-space vertical  force &$\frac{-f^{(q)} x \log \left(q^2+x^2\right)}{4 \pi \mu}$ & $\frac{ f^{(q)}  q}{\pi  \mu \left(q^2+x^2\right)}$\\ \hline
  Isotropic half-space horizontal  force &$\frac{f^{(x)} q \left(\log \left(q^2+x^2\right)+2\right)}{4 \pi\mu }$ & $\frac{ f^{(x)} x}{\pi \mu  \left(q^2+x^2\right)}$\\ \hline
   Pre-strained half-space vertical  force &$ \frac{-f^{(q)}  \lambda^2 }{2 \mu \pi  \left(\lambda^8+2 \lambda^4-4 \lambda^2+1\right)}\left(2 q \left(\lambda^4+1\right) \tan ^{-1}\left(\frac{x}{q}\right)-4 q \lambda^2 \tan ^{-1}\left(\frac{x}{q \lambda^2}\right)\right.$ $\left.+x \left(\lambda^4+1\right) \log \left(q^2+x^2\right)-2 x \log \left(q^2 \lambda^4+x^2\right)\right)$ & $\frac{f^{(q)}  q \left(1+\lambda^2\right) \left(1+\lambda^4\right)}{\pi \mu \left(q^2+x^2\right) \left(-1+3 \lambda^2+\lambda^4+\lambda^6\right)}$\\ \hline
  Pre-strained half-space horizontal  force &$\frac{-f^{(x)}  \lambda^2}{2 \mu \pi  \left(\lambda^8+2 \lambda^4-4 \lambda^2+1\right)} \left(4 x \lambda^2 \tan ^{-1}\left(\frac{q}{x}\right)-2 x \left(\lambda^4+1\right) \tan ^{-1}\left(\frac{q \lambda^2}{x}\right)\right.$ $\left.-q \lambda^2 \left(\left(\lambda^4+1\right) \log \left(q^2 \lambda^4+x^2\right)-2 \log \left(q^2+x^2\right)\right) \right.$ $\left.+x\pi \left(-1+\lambda ^2\right)^2 (1-2 \theta(x))\right)$ & $\frac{2 f^{(x)} x \lambda^2 \left(1+\lambda^2\right)}{\pi \mu  \left(q^2+x^2\right) \left(-1+3 \lambda^2+\lambda^4+\lambda^6\right)}$\\ \hline
    \end{tabular}
\end{center}
\caption{Half space surface force Green's functions}\label{surfacetab}
\end{table}

\section{Exact volume conservation for three-dimensional axisymmetric deformations}
\begin{figure}[h]\centering
\includegraphics[width=0.4 \textwidth]{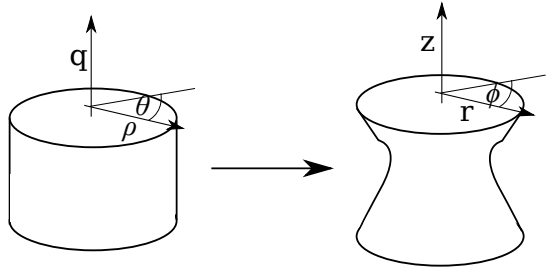}\caption{A three dimensional axisymmetric elastic body (left) labeled by the coordinates $(\rho,\theta,q)$ is deformed into the target state (right) labeled by the coordinates $(r,\phi,z)$.}\label{targandrefaxi}
\end{figure}
We now turn to 3d axisymmetric deformations of incompressible materials. We consider a neo-Hookean elastic body labeled by the  polar-coordinate system $(\rho,\theta, q)$ in the reference state that is deformed into a target or current state parameterized by $(r, \phi,z)$, depicted in fig.\ \ref{targandrefaxi}. We usually describe an axisymmetric deformation with functions $r(\rho,q)$ and $z(\rho,q)$, with axisymmetry requiring that $\phi=\theta$ for each point in the elastic body. Within this description, in a polar coordinate system, the deformation gradient tensor is 
\begin{equation}
F_1(\rho,q)=\left(\begin{array}{ccc} 
\frac{\partial r}{\partial  \rho}\big|_q&0&\frac{\partial  r}{\partial  q}\big|_\rho\\\
0&r/\rho&0\\
\frac{\partial  z}{\partial  \rho}\big|_q&0&\frac{\partial  z}{\partial  q}\big|_\rho\end{array}\right).
\end{equation}
Once again the constraint of volume conservation requires that ${\rm det} F_1=1$. As in the 2-D case, we include the possibility that, in the $\rho-q$ elastic reference state the material has already undergone an axisymmetric and volume preserving pre-strain $F_0=\mathrm{diag}(\lambda,\lambda,1/\lambda^2)$, so that the total deformation from the unstrained state is $F_1\cdot F_0$. Again, as in the 2-D case, the equations of equilibrium are then 
\begin{equation}
\nabla \cdot (\mu F_1\cdot F_0\cdot{F_0}^{T}-P F_1^{-T})=-\mathbf{f} \hspace{3 em} \Det{F_1}=1,\label{tradnleqns2}
\end{equation}
with the divergence being taken in the  $(\rho,\theta,q)$ reference state. Inspired by the Gaussian incompressible mapping in the two dimensional case, we choose to instead represent the deformation $F_1$ via the functions $\rho(r,q)$  and $z(r,q)$ that reside partly in the reference and partly in the current configurations, so that $F_1$ becomes
\begin{equation}
F_1(r,q)=\left(\begin{array}{ccc} 
\left(  \frac{\partial \rho}{\partial r}\big|_q\right)^{-1}&0& -\frac{\partial \rho}{\partial q}\big|_r \left(\frac{\partial \rho}{\partial r }\big|_q\right)^{-1} \\\
0&r/\rho&0\\
 \frac{ \partial z}{\partial r}\big|_q \left(\frac{\partial \rho}{\partial r }\big|_q\right)^{-1} &0&\frac{\partial z}{\partial q}\big|_r-\frac{\partial \rho}{\partial q}\big|_r \frac{\partial z}{\partial r}\big|_q\left(\frac{\partial \rho}{\partial r}\big|_q\right)^{-1}\end{array}\right).
\end{equation}
Assuming all quantities are functions of $r$ and $q$, so that $\rho_r=\frac{\partial \rho}{\partial r}\big|_q$, we can calculate 
\begin{equation}
\det{F_1}=\frac{r z_q}{\rho \rho_r}.
\end{equation}
To enforce perfect volume conservation, we introduce the scalar field $\chi(r,q)$  defined by the relations
\begin{equation}
z(r,q)=\frac{\chi_r}{r} \mathrm{\ \ \ \ \ and \ \ \ \ \ \ }\rho(r,q)= \sqrt{2 \chi_q},
\end{equation}
so that $\det{F}=\slfrac{\left(r \frac{1}{ r}\chi_{rq}\right)}{\left( \rho (2 \chi_q)^{-1/2}\chi_{rq}\right)}=1.$ In terms of this new  field, the deformation gradient is
\begin{equation}
F_1(r,q)=\left(
\begin{array}{ccc}
 \frac{ \sqrt{2\chi _q}}{\chi _{{rq}}} & 0 & -\frac{\chi _{{qq}}}{\chi _{{rq}}} \\
 0 & \frac{r}{2 \sqrt{\chi _q}} & 0 \\
 \frac{\sqrt{2\chi _q} \left(-\chi _r+r \chi _{{rr}}\right)}{r^2 \chi _{{rq}}} & 0 & \frac{r \chi _{{rq}}^2+\chi _{{qq}} \left(\chi _r-r \chi _{{rr}}\right)}{ r^2 \chi _{{rq}}}
\end{array}
\right).\label{axiSF}
\end{equation}

\subsection{Incremental axisymmetric three dimensional elasticity}
Before any additional displacement ($F_1=I$) we have $\chi=\half r^2 q$ and $P=\mu P_0$, where $\mu P_0$ may be a large pressure associated with the pre-strain. To linearize about this reference state, we write
\begin{equation}
\chi(r,q)=\half r^2 q+\beta(r,q) \mathrm{,\ \ \ \ and\ \ \ \ }P(r,q)=\mu P_0+\mu\kappa(r,q),
\end{equation}
where $\beta,\mathrm{\ } \kappa<<1$. Expanding $F$ and $F^{-T}$ to linear order in $\beta$ give
\begin{align}
F= \left(
\begin{array}{ccc}
 1+\frac{\beta_{q}-r \beta_{qq}}{r^2} & 0 & -\frac{\beta_{qq}}{r} \\
 0 & 1-\frac{\beta_{q}}{r^2} & 0 \\
 \frac{r \beta_{rr}-\beta_{r}}{r^2} & 0 &1+ \frac{\beta_{qq}}{r} \\
\end{array}
\right)\hspace{3em} F^{-T}=\left(
\begin{array}{ccc}
 1-\frac{\beta_{q}-r \beta_{qq}}{r^2}& 0 & \frac{\beta_{r}-r \beta_{rr}}{r^2} \\
 0 & 1+\frac{\beta_{q}}{r^2} & 0 \\
 \frac{\beta_{qq}}{r} & 0 & 1-\frac{\beta_{qq}}{r} \\
\end{array}
\right).
\end{align}
As claimed in our letter, these forms are algebriacally identical to those that would be derived by introducing a traditional Stokes stream line function $\beta(\rho,q)$ such that $r=\rho-(1/\rho)\beta_q$ and $z=q+(1/r) \beta_\rho$, with the identification $\rho\to r$. We then find the linearized equations of equilibrium by substituting these results into eqn.\ (\ref{tradnleqns2}), and expanding to first order. To conduct the expansion we must recall the form for the divergence of a tensor in cylindrical polars (see, for example \cite{bower2011applied} Appendix D) and make use of the first order partial derivative identities  $\frac{\partial}{\partial \rho}\big|_q=\frac{\partial}{\partial r}\big|_q$ and $\frac{\partial}{\partial q}\big|_\rho=\frac{\partial}{\partial q}\big|_r$, to get
  \begin{equation}
\frac{\mu}{\lambda ^4 r^2}\left(
\begin{array}{c}
 - r^2 \lambda^4 \kappa_r-r   \beta _{{qqq}}+\lambda^6   \left(\beta _{{rq}}-r \beta _{{rrq}}\right)\\
 - r^2 \lambda^4 \kappa_q+r \beta _{{rqq}}+\lambda^6 \left(r \beta _{{rrr}}-\beta _{{rr}}+\beta_r/r\right) 
 \end{array}
\right)=-\mathbf{f}.\label{axianisoequibeqns}
\end{equation}
In a region with no external force ($\mathbf{f}=0$) we can once again eliminate $\kappa$ to get the axsysmetric version of the 2D eqn.\ (\ref{mastereqn2d}),
\begin{equation}
\left(\lambda^6 r\frac{\partial}{\partial r}\left(\frac{1}{r}\frac{\partial }{\partial r}\right)+\frac{\partial^2}{\partial q^2}\right)\left(r\frac{\partial}{\partial r}\left(\frac{1}{r}\frac{\partial }{\partial r}\right)+\frac{\partial^2}{\partial q^2}\right)\beta(r,q)=0\label{axianisogoverningequation}.
\end{equation}
In the case where there is no pre-strain ($\lambda=1$) these two equations reduce to
\begin{equation}
\frac{\mu}{ r^2}\left(
\begin{array}{c}
  \beta _{{rq}}- r(r  \kappa_r+   \beta _{{qqq}}+ \beta _{{rrq}})  \\
 - r^2 \kappa_q+r (\beta _{{rqq}}+  \beta _{{rrr}})-\beta _{{rr}}+\beta_r/r
 \end{array}
\right)=-\mathbf{f}.\label{axiequib}
\end{equation}
and an axisymmetric analog of the biharmonic equation discussed in our letter,
\begin{equation}
\left(r\frac{\partial}{\partial r}\left(\frac{1}{r}\frac{\partial }{\partial r}\right)+\frac{\partial^2}{\partial q^2}\right)^2\beta=0\label{axiisogoverningequation}.
\end{equation}

\subsection{Axisymmetric  Green's functions for unstrained full space}
In three dimensions, for a force along the axis of symmetry, we expect the displacement to vary inversely as  the distance from the point of application of the force, so $\beta$ should increase proportional to the distance from the point of application, leading us to the  suggestion that
\begin{equation}
\beta(r,q)=A \frac{r^2}{(q^2+r^2)^{1/2}}+B \frac{q^2}{(q^2+r^2)^{1/2}}.
\end{equation}
This function satisfies the equation of equilibrium (eqn.\ (\ref{axiisogoverningequation})), but axisymmetry requires that $\rho(0,q)=0$, so $B=0$. Substituting $\beta$ into eqn.\ (\ref{axiequib}), we find that  $\kappa(r,q)=2A q\left(q^2+r^2\right)^{-3/2}$. Normalizing this function so that it is the response to a total force $f$, the Green's functions are
\begin{equation}
\beta(r,q)=\frac{f r^2}{8 \pi  \mu  \sqrt{q^2+r^2}}\mathrm{\ \ \ \ and \ \ \ \ }\kappa(r,q)=\frac{f q}{4 \pi \mu \left(q^2+r^2\right)^{3/2}  },
\end{equation}
which, at linear-order, match the elementary Kelvin solution \citep{kelvin1848displacement} in linear elasticity.

\begin{figure}\centering
\subfigure[Inward ring-load.]{\label{streamline}\includegraphics[width=0.33\textwidth]{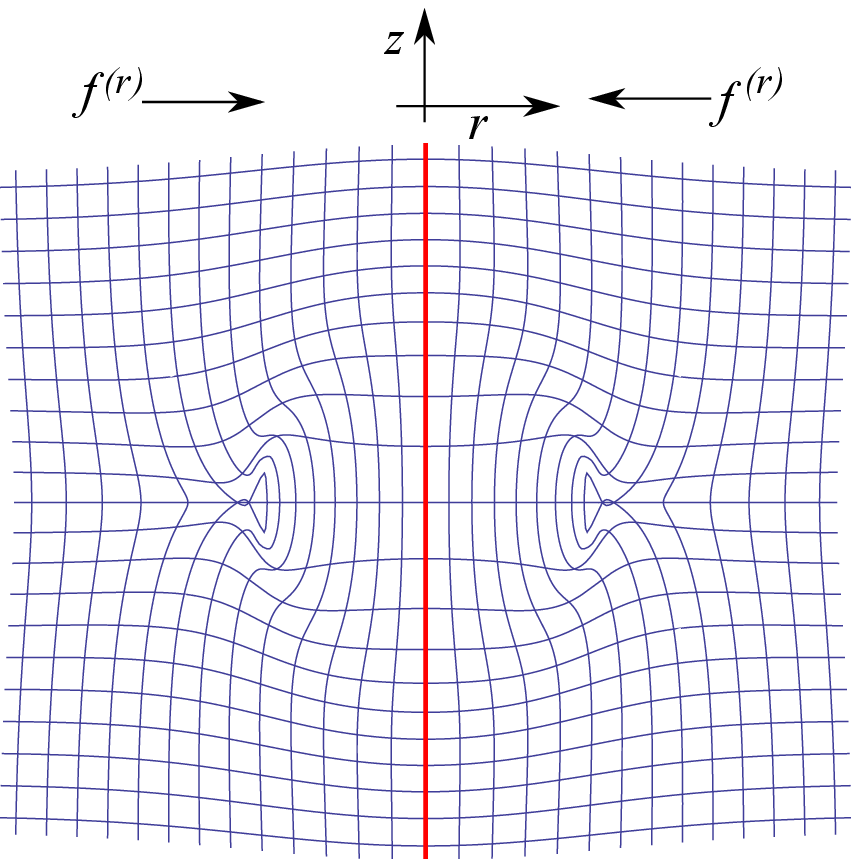}}
\subfigure[Upward ring load.]{\label{linear}\includegraphics[width=0.33\textwidth]{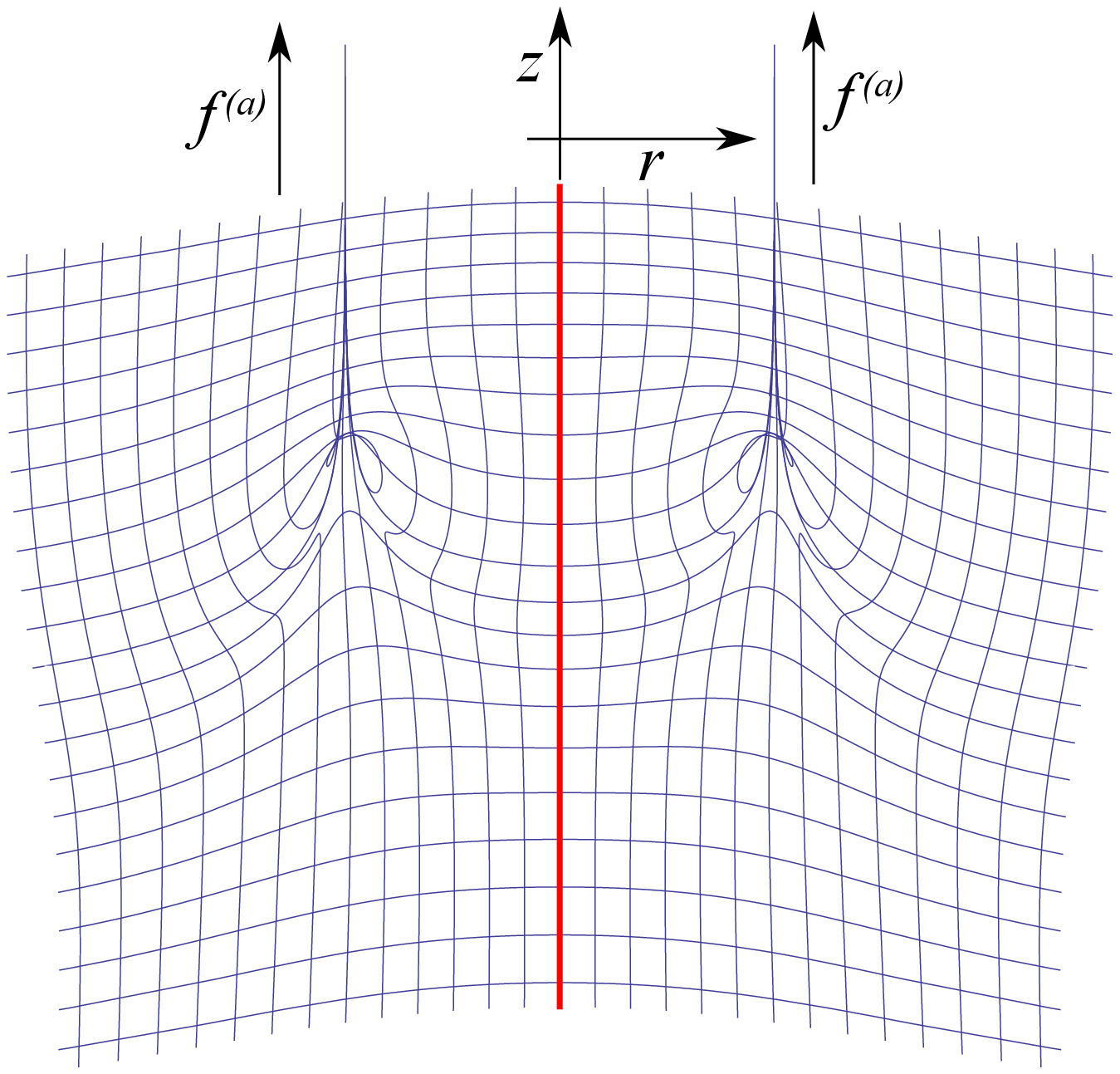}}
\caption{Exactly volume preserving deformation caused by an inward radial ring load (left, eqn.\ (\ref{radialring})) and an upward radial ring load (right, eqn.\ (\ref{axialring})). The red line indicates the axis of symmetry, the grid was a square lattice with spacing $0.1$ in the reference state, the total force applied in both cases is $2 \pi \mu$, and the ring is located at $r=0.55$.}\label{axisymgreens}
\end{figure}

Since we are restricted to axisymmetric situations we cannot consider point-forces in other directions acting on the axis, or point forces in any direction acting away from the axis. However, we can consider both radial and axial forces acting in rings around the axis. In these cases there are no simple scaling arguments that can be used to produce the solutions. In conventional linear elasticity the Green's function for a ring-load can be found by using the elastic reciprocal theorem \citep{kermanidis1975numerical} or integrating the point-force solution around a ring \citep{hanson1997concentrated}. Here, we find these Green's functions by taking these traditional results, expressing them (for the incompressible case) via the traditional Stokes streamline function then identifying $\rho\to r$.  For a ring load applying a force along the axis, we get
\begin{align}
\beta^{(Ra)}(r,q)&=\frac{f^{(a)}}{8 \pi ^2 \mu }\sqrt{(a+r)^2+q^2} \left(\left(\frac{2 r (a+r)}{(a+r)^2+q^2}-1\right) K\left(\frac{4 a r}{q^2+(a+r)^2}\right)+E\left(\frac{4 a r}{q^2+(a+r)^2}\right)\right)\label{axialring}\\
\kappa^{(Ra)}(r,q)&=\frac{f^{(a)}}{4 \pi ^2 \mu }\frac{2 q E\left(\frac{4 a r}{q^2+(a+r)^2}\right)}{\left((a-r)^2+q^2\right) \sqrt{(a+r)^2+q^2}},
\end{align}
where $f^{(a)}$ is the total force in the applied in the axial direction along a ring of radius $a$ and $K(m)=\int_0^{\pi/2}\frac{\mathrm{d}\theta}{\sqrt{1-m \sin^2{\theta}}}$ and $E(m)=\int_0^{\pi/2}\sqrt{1-m \sin^2{\theta}}\mathrm{d}\theta$ are the complete elliptic integrals of the first and second kind.   Similarly, if the applied force is in the radial direction along the same ring, the Green's function are:
\begin{align}
\beta^{(Rr)}(r,q)&=\frac{f^{(r)} q \left(\left((a+r)^2+q^2\right) E\left(\frac{4 a r}{q^2+(a+r)^2}\right)-\left(a^2+q^2+r^2\right) K\left(\frac{4 a r}{q^2+(a+r)^2}\right)\right)}{8 \pi^2 a \mu   \sqrt{(a+r)^2+q^2}}\label{radialring}\\
\kappa^{(Rr)}(r,q)&=\frac{f^{(r)} \left(\left(-a^2+q^2+r^2\right) E\left(\frac{4 a r}{q^2+(a+r)^2}\right)-\left((a-r)^2+q^2\right) K\left(\frac{4 a r}{q^2+(a+r)^2}\right)\right)}{4 \pi^2 a \mu  \left((a-r)^2+q^2\right) \sqrt{(a+r)^2+q^2}}\label{radialringkappa}.
\end{align}
These two solutions are plotted in fig. \ref{axisymgreens}. As before there is some self intersection near the point of application of the force caused by the divergent stress, where the solution is expected to break down.

\subsection{Axisymmetric Green's functions for pre-strained full space}
As in the case without the pre-strain, we expect $\beta(r,q)$ to be proportional to distance from the point of application of the force. Taking inspiration from the analogous 2-D case, the simplest functions with this scaling at large distances which satisfy the equation (\ref{axianisogoverningequation}) for $\beta$ are
$\beta_1=\sqrt{r^2+q^2},$ and $\beta_2=\sqrt{r^2+\lambda^6 q^2}$. Substituting these forms into eqn. (\ref{axianisoequibeqns}), we see the incremental pressures associated with them are
\begin{equation}
\kappa_1(r,q)=\frac{q \left(-1+\lambda^6\right) }{\left(q^2+r^2\right)^{3/2}\lambda^4} \mathrm{\ \ \ \ and \ \ \ \ }\kappa_2(r,q)=0.
\end{equation}
However, as in the 2-dimensional case, these are not admissible deformations because  $\rho(0,q)\ne0$. As in the 2-D case, we can take a linear combination of $\beta_1$ and $\beta_2$ that is admissible:
\begin{equation}
\beta^{(a)}(r,q)=\frac{f \lambda}{4 \pi \mu  \left(\lambda^6-1\right)  }\left(\lambda^3\sqrt{q^2+r^2}- \sqrt{r^2+q^2 \lambda^6}\right)\label{betaaxiwholespace}.
\end{equation}
Taking the same linear combination of $\kappa_1$ and $\kappa_2$ we see the incremental pressure field is given by
\begin{equation}
\kappa^{(a)}(r,q)=\frac{f q}{4 \pi  \mu \left(q^2+r^2\right)^{3/2}  }\label{kappaaxiwholespace}.
\end{equation}
The pre factor has been chosen by integration of the stress over an infinite cylinder oriented along the axis of symmetry. The ring-load Green's functions in the presence of a large pre-stress are not expressible in closed form, so they will not be presented here.

\subsection{Green's functions for half spaces with large pre-strains}
We find the on-axis half-space Green's function for an axially symmetric  point force at $(0,-d)$ in an analogous way to the planar problem treated earlier. The bulk equations of equilibrium (eqns.\ (\ref{axianisoequibeqns})) are now supplemented by the condition that the free surface at $q=0$ be stress free, which, from eqn.\ (\ref{tradnleqns2}), we see reads
\begin{equation}
(\mu F_1\cdot F_0\cdot F_0^{T}-P F_1^{-T})\cdot(0,0,1)=(0,0,0).\label{surfacestressaxi}	
\end{equation}
Linearizing this with respect to $\beta$ and $\kappa$ yields
\begin{equation}
\frac{\mu }{ r^2 \lambda ^4}\left(\lambda ^4 P_0 \left(r \beta_{rr}(r,0)-\beta_{r}(r,0)\right)-r \beta_{qq}(r,0),0,r \left(\left(\lambda ^4 P_0+1\right) \beta_{qq}(r,0)-\lambda ^4 r \kappa (r,0)\right)+r^2 \left(1-\lambda ^4 P_0\right)\right)=\left(0,0,0\right)
.\end{equation}
Setting $\beta\to0$ and $\kappa \to 0$, we see that $P_0=1/\lambda^4$. Inspired by the solution in the 2-D place, we then try for the 3-D half space solution  the whole space Green's function (eqns.\ (\ref{betaaxiwholespace}) and (\ref{kappaaxiwholespace})) at a depth $d$ below the free surface augmented by image forces above the half space, 
 in the form
\begin{align}
\beta(x,q)&=\beta^{(a)}(r,q+d)+a_1 \beta_1(r,q-d)+a_2 \beta_2(r,q-d)+a_3 \beta_1(r,q-d \lambda^3)+a_4 \beta_2(r,q-d/ \lambda^3)\notag \\
\kappa(x,q)&=\kappa^{(a)}(r,q+d)+a_1 \kappa_1(r,q-d)+a_2 \kappa_2(r,q-d)+a_3 \kappa_1(r,q-d \lambda^3)+a_4 \kappa_2(r,q-d/ \lambda^3)\label{axifullspaceanisogreensfunc},
\end{align}
analogous to eqn.\ (\ref{halfspacverticalnoniso}). We can then satisfy the boundary condition with the choices:
\begin{align}
a_1=\frac{f \lambda^4 \left(1+3 \lambda^3-\lambda^6+\lambda^9\right)}{4 \pi  \left(-1+\lambda^3\right)^2 \left(-1+3 \lambda^3+\lambda^6+\lambda^9\right) \mu }\mathrm{,\ }a_3&=\frac{- f \lambda^4 \left(1+\lambda^6\right)}{\pi  \left(-1+\lambda^3\right)^2 \left(1+\lambda^3\right) \left(-1+3 \lambda^3+\lambda^6+\lambda^9\right) \mu } \notag \\ 
a_2= a_1/\lambda^3\mathrm{,\ \ \ \ \ \ \ \  \ \ \ \ \ \ \ \ \ \ \ \ \ \   \ \ \ \ \ \ \  \ \ \ \ \ \ \ \ \ \ \ \ \ \ \ \ }a_4&=a_3.\label{a1axicoef}
\end{align}

As in the two dimensional case, although these fields satisfy the equations of mechanical equilibrium (eqn.\ (\ref{axianisoequibeqns})) and the boundary condition (eqn.\ (\ref{surfacestressaxi})), they are not admissible because they do not have continuous  displacements since $\rho(0,q)\ne 0$. However, we can fix this by adding $\frac{q f \lambda^4 \left(-1+\lambda^3\right)}{2\pi  \left(-1+3 \lambda^3+\lambda^6+\lambda^9\right) \mu }$ to $\beta$ given in eqn. (\ref{axifullspaceanisogreensfunc}), which does not give rise to any (first order) stress or strain but removes the discontinuity at the origin. No modification to $\kappa$ is required.

\subsection{Axisymmetric Green's functions for unstrained half space}
Taking the isotropic limit ($\lambda\to 1$) of eqn.\ (\ref{axifullspaceanisogreensfunc})  gives the point force half space Green's function as
\begin{equation}
\beta(r,q)=\frac{f r^2}{8 \pi  \mu }\left(\frac{1}{\sqrt{(d-q)^2+r^2}}+\frac{1}{\sqrt{(d+q)^2+r^2}}-\frac{2 d q}{\left((d-q)^2+r^2\right)^{3/2}}\right)
\end{equation}
\begin{equation}
\kappa(r,q)=\frac{f}{4 \pi \mu }\left(\frac{d+q}{\left((d+q)^2+r^2\right)^{3/2}}-\frac{(d-q)^2 (5 d-q)-(d+q) r^2}{\left((d-q)^2+r^2\right)^{5/2}}\right),
\end{equation}
which matches Mindlin's solution \citep{mindlin1936force} to linear order. Taking  $d \to 0$  gives the surface Green's functions, corresponding to those found by Boussinesq \citep{boussinesq1885application} and Cerruti  \citep{cerruti1893sulla}.

Isotropic half-space  ring-load Green's functions are known in conventional linear elasticity \citep{hanson1997concentrated}. Ours will be the same combination of image forces. For an axial ring loading this is:
\begin{align}
\beta(r,q)&=\beta^{(Ra)}(r,q+d)+\beta^{(Ra)}(r,q-d)+\frac{2 d q}{q-d}\beta^{(Ra)}_q(r,q-d)\label{axialhalfspace}\\
\kappa(r,q)&=\kappa^{(Ra)}(r,q+d)+\kappa^{(Ra)}(r,q-d)+2 d \kappa^{(Ra)}_q(r,q-d),
\end{align}
while for a radial ring loading we have:
\begin{align}
\beta(r,q)&=\beta^{(Rr)}(r,q+d)+\beta^{(Rr)}(r,q-d)+\frac{2 d (2 q-d) }{(d-q)^2}\beta^{(Rr)}(r,q-d)+\frac{2 d q }{d-q}\beta^{(Rr)}_q(r,q-d)\notag \\
\kappa(r,q)&=\kappa^{(Rr)}(r,q+d)+\kappa^{(Rr)}(r,q-d)-2 d \kappa^{(Rr)}_q(r,q-d),\label{radialhalfspace}
\end{align}
where $\beta^{(Rr)}$ etc.\ are the full space ring Green's functions given in eqns.\ (\ref{axialring}-\ref{radialringkappa}).

\begin{figure}\centering
\subfigure[Half-space outward ring-load.]{\label{streamline}\includegraphics[width=0.32\textwidth]{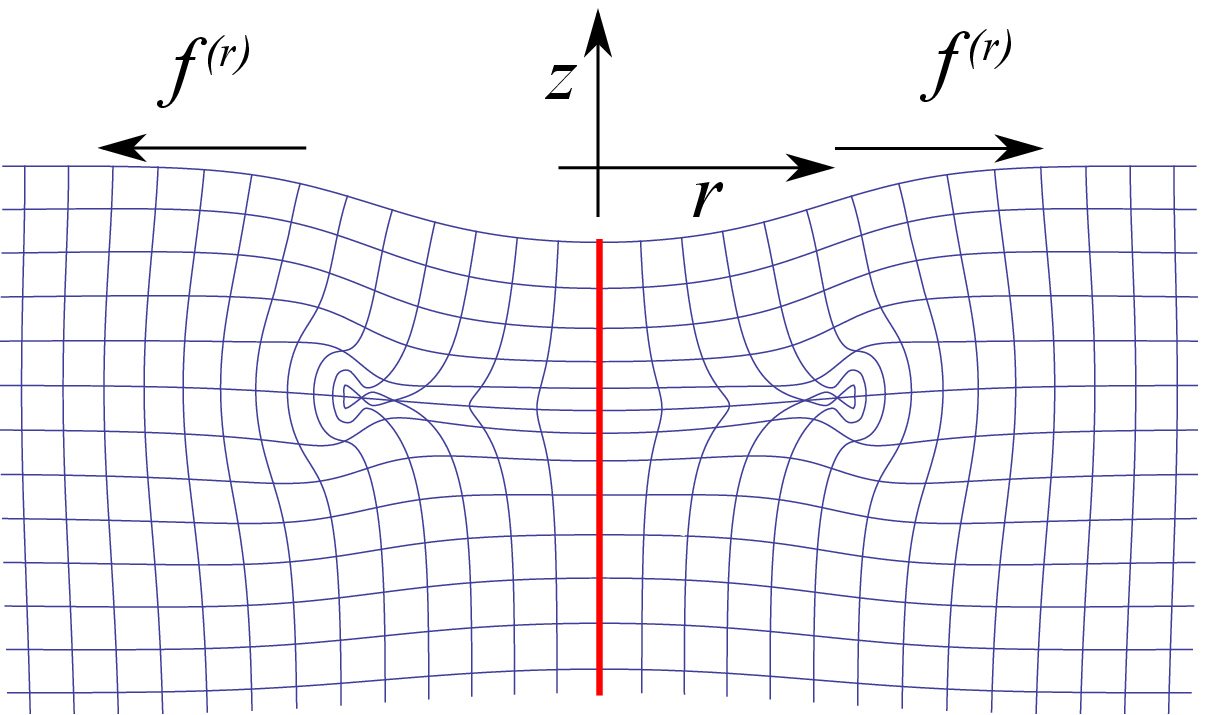}}
\subfigure[Half-space inward ring-load.]{\label{streamline}\includegraphics[width=0.32\textwidth]{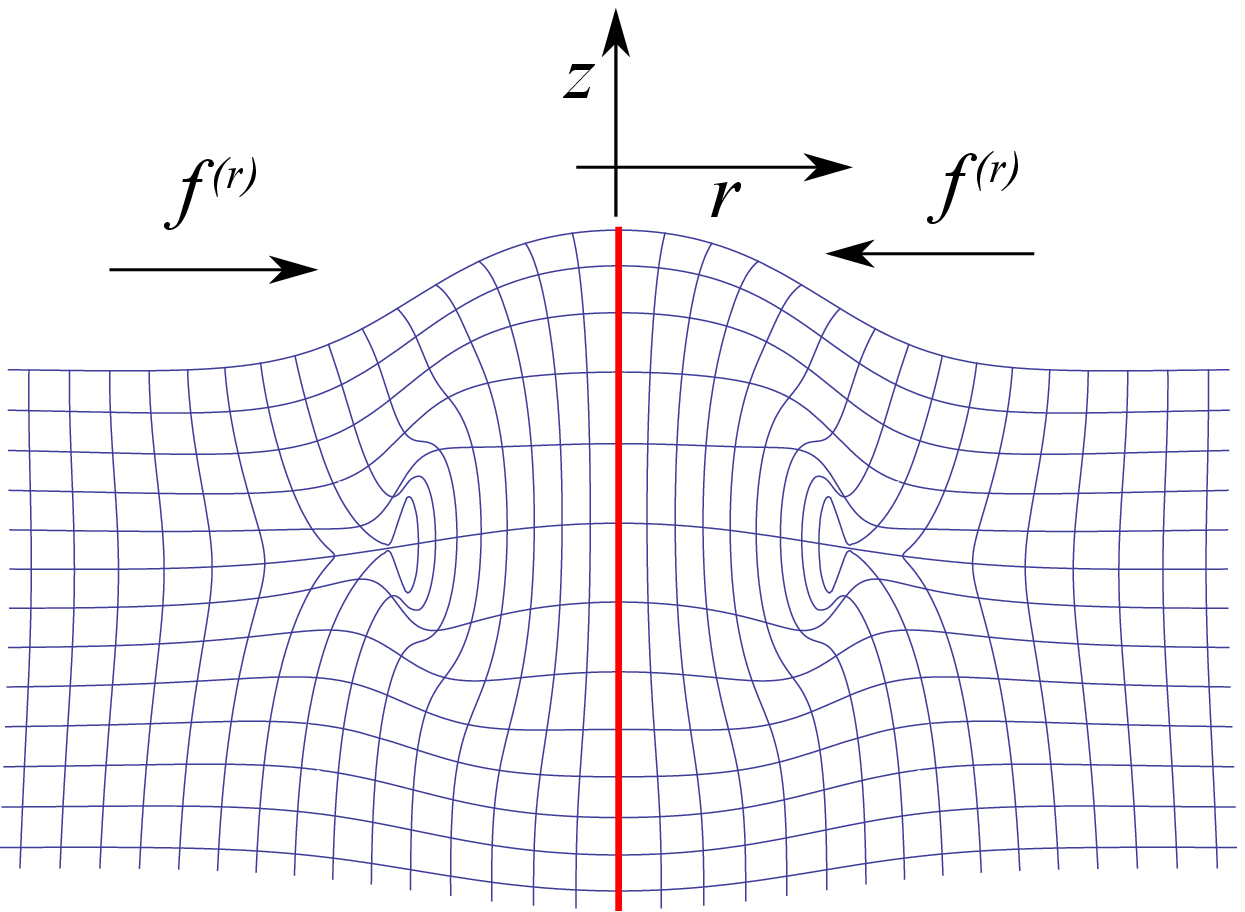}}
\subfigure[Half-space upward ring load.]{\label{linear}\includegraphics[width=0.32\textwidth]{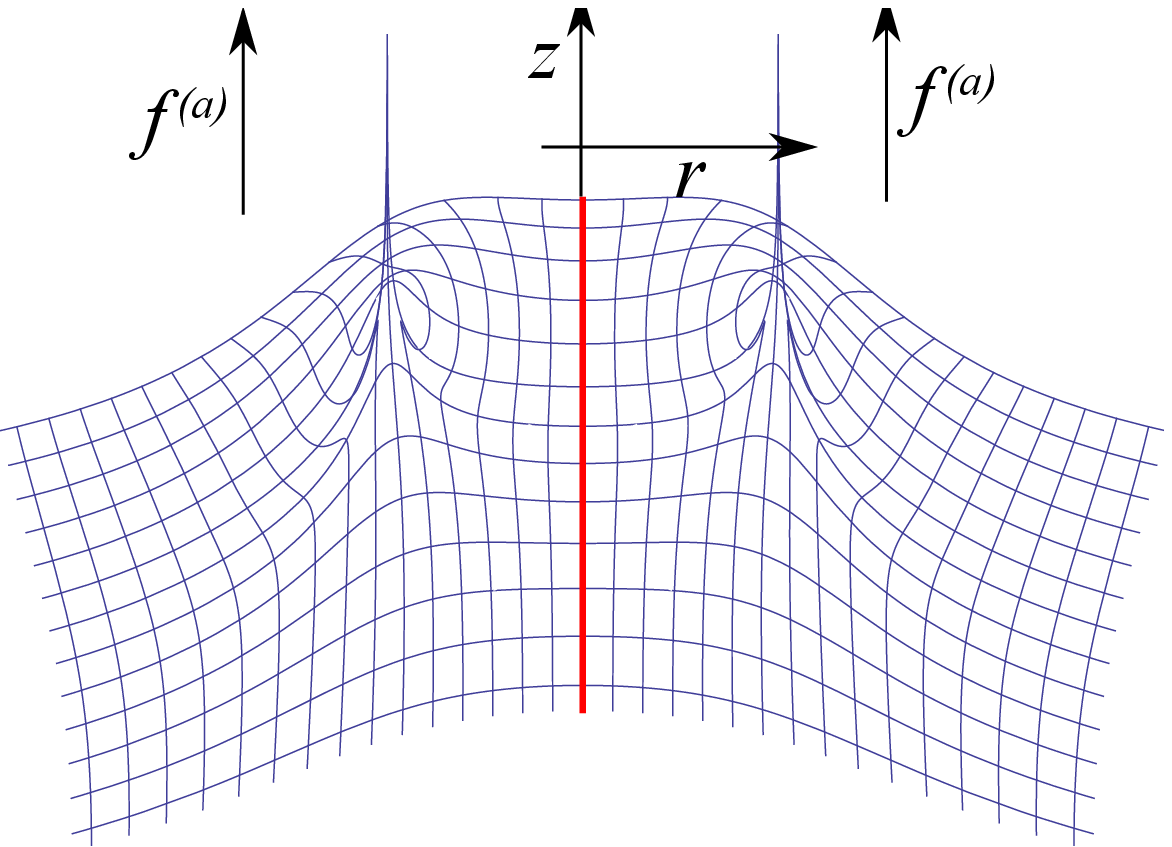}}
\caption{Deformation caused by outward and inward radial ring loads (subfigures a and b, eqn. (\ref{radialhalfspace}) and an upward ring load (right, eqn.\ (\ref{axialhalfspace})) in an axisymmetric half-space. The red line indicates the axis of symmetry, the grid was a square lattice with spacing $0.1$ in the reference state and the ring is located at $r=0.55$. The force applied was $2 \pi \mu$ for the inward and upward loadings and  $\pi \mu$ for the outward loading.}\label{fig:halfspaceoffaxisplots}

\end{figure}

As in the full-space case, it is simple but algebraically very laborious to show that these fields satisfy the bulk and boundary equations  (but see Supplementary Mathematica notebooks). These solutions are plotted in fig.\ \ref{fig:halfspaceoffaxisplots}. We find that the strains are higher than those in the full space case (fig.\ \ref{axisymgreens}).

\section{Surface instability deduced from half space Green's functions}
As first shown by Biot \citep{biotbook} the surface of compressed elastic half spaces become unstable to the formation of creases  at a critical large compression, although recent results have shown that before this instability is reached, there is a sub-critical instability with no nucleation threshold \citep{hohlfeld2011, hohlfeld2012}. While the nonlinear instability can not be deduced from a linear calculation as here, it is worth mentioning that the original Biot instability manifests itself in the half space Green's functions for pre-strained solids. This is most clearly seen in the dependence of the functions via the pre-strain $\lambda$ in the image-charge potentials. At the point of instability the dependence lead to a diverging response and a reversal in the 
sign of the displacement caused by a force. In the two-dimensional case, the response of the half-space diverges when the denominator in eqn.\ (\ref{a1coef}) vanishes, which is when
$-1+3 \lambda^2+\lambda^4+\lambda^6=0$, so that the instability occurs when 
\begin{equation}
\lambda^*=\frac{1}{3} \left(-1-2\left(17+3 \sqrt{33}\right)^{-1/3}+\left(17+3 \sqrt{33}\right)^{1/3}\right)\approx0.543689...
\end{equation}
The denominator also vanishes when $\lambda=1$, but in this case the images associated with $a_1$ and $a_3$  collapse onto the same point and cancel out, so there is no divergent response. Similarly, in the  three-dimensional axisymmetric case, the onset of instability occurs when the denominator in eqn.\ (\ref{a1axicoef}) vanishes, which occurs when
$-1+3 \lambda^3+ \lambda^6+\lambda^{9}=0$,
giving
\begin{equation}
\lambda^*=\left(\frac{1}{3}\left(-1-4\times 2^{2/3}\left(13+3 \sqrt{33}\right)^{-1/3}+\left(26+6 \sqrt{33}\right)^{1/3}\right)\right)^{1/3}\approx0.666142... 
\end{equation}
These thresholds agree with those originally found by Biot \citep{biotbook}. This is expected in the planar case, but is perhaps surprising in the axisymmetric case where Biot included an axisymmetric pre-strain, but only accounted for two dimensional plane-strain perturbations. A recent numerical study of the axisymmetric Biot problem confirms this result \citep{Tallinen2013}. Our results are for the simplest possible non-linear elastic constitutive relation (neo-Hookean) but the surface instability is found in a very wide class of materials \citep{brun2003boundary}.

%

\end{document}